\documentclass[numberedappendix]{emulateapj}

\usepackage{amsmath}
\usepackage{amssymb}
\usepackage{graphicx}

\slugcomment{Submitted manuscript}
\shorttitle{Kinetic Alfv\'en and slow waves}
\shortauthors{Zhao et al.}

\begin{document}

\title{Properties of Short-wavelength Oblique Alfv\'{e}n and Slow Waves}
\author{J. S. Zhao$^{1,2,3}$ \and Y. Voitenko$^4$ \and M.~Y.~Yu$^{5,6}$ \and J. Y.
Lu$^7$ \and D. J. Wu$^1$}

\begin{abstract}
Linear properties of kinetic Alfv\'{e}n waves (KAWs) and kinetic slow waves
(KSWs) are studied in the framework of two-fluid magnetohydrodynamics. We
obtain the wave dispersion relations that are valid in a wide range of the
wave frequency $\omega $ and plasma-to-magnetic pressure ratio $\beta $. The
KAW frequency can reach and exceed the ion cyclotron frequency at ion
kinetic scales, whereas the KSW frequency remains sub-cyclotron. At $\beta
\sim 1$, the plasma and magnetic pressure perturbations of both modes are in
anti-phase, so that there is nearly no total pressure perturbations.
However, these modes exhibit also several opposite properties. At high $%
\beta $, the electric polarization ratios of KAWs and KSWs are opposite at
the ion gyroradius scale, where KAWs are polarized in sense of electron
gyration (right-hand polarized) and KSWs are left-hand polarized. The
magnetic helicity $\sigma \sim 1$ for KAWs and $\sigma \sim -1$ for KSWs,
and the ion Alfv\'{e}n ratio $R_{Ai}\ll 1$ for KAWs and $R_{Ai}\gg 1$ for
KSWs. We also found transition wavenumbers where KAWs change their
polarization from left- to right-hand. These new properties can be used to
discriminate KAWs and KSWs when interpreting kinetic-scale electromagnetic
fluctuations observed in various solar-terrestrial plasmas. This concerns,
in particular, identification of modes responsible for kinetic-scale
pressure-balanced fluctuations and turbulence in the solar wind.
\end{abstract}


\affil{1 Purple Mountain Observatory, Chinese Academy of Sciences, Nanjing 210008,
China; js\underline{ }zhao@pmo.ac.cn}
\affil{2 Key Laboratory of Solar activity, National Astronomical
Observatories, Chinese Academy of Sciences, Beijing 100012, China.}
\affil{3 Key Laboratory of Modern Astronomy and Astrophysics, Nanjing
University, Nanjing 210093, China.}

\affil{4 Solar-Terrestrial Centre of Excellence, Space Physics Division,
Belgian Institute for Space Aeronomy, Avenue Circulaire 3, B-1180 Brussels,
Belgium}

\affil{5 Institute for Fusion Theory and Simulation and Department of Physics,
Zhejiang University, Hangzhou 310027, China}
\affil{6 Institute for
Theoretical Physics I, Ruhr University, D-44780 Bochum, Germany}

\affil{7 College of Math and Statistics, Nanjing University of Information Science and Technology,
Nanjing 210044, China.}

\affil{Purple Mountain Observatory, Chinese Academy of Sciences, Nanjing,
China}

\section{Introduction}

Kinetic Alfv\'{e}n waves (KAWs) have been receiving recently much attention
in connection to understanding turbulence at kinetic scales in the solar
wind and the near-Earth space environment \cite{CH08,PO13}. KAWs can be
generated by the MHD Alfv\'{e}nic turbulence through an anisotropic cascade
\cite{HO08,BI10,ZH13}, or through non-local coupling of the MHD Alfv\'{e}n
waves \cite{ZH11,ZH14}. These processes provide a pathway for the turbulence
to dissipate via KAWs' damping \cite{SC09}. KAWs have been found in many in
situ spacecraft measurements \cite{CH08,CH09,HU12,PO13}. Identification of
KAWs is usually accomplished by analyzing characteristic wave parameters,
such as the ratio of electric to magnetic perturbations \cite{CH08,CH09},
magnetic compressibility \cite{PO12}, magnetic helicity \cite{HO10,PO11,HE12}%
, or the profile of wave dispersion \cite{SA09,SA10,RO13}.

On the other hand, kinetic slow waves (KSWs) have been found indirectly by
analyzing the compressible turbulent fluctuations in the solar wind
turbulence \cite{HO12,KL12}. KSWs have been used in the interpretation of
recent observations of small-scale pressure balanced structures (PBSs) in
the solar wind \cite{YAO11,YAO13}, which exhibit an anti-correlation between
the plasma and the magnetic pressure fluctuations. Since such small-scale
PBSs can also be associated with KAWs, it is of great interest to
investigate in more detail properties of the KAW and KSW modes at parallel
and perpendicular kinetic scales, especially their differences.

Hollweg (1999) derived a two-fluid dispersion equation for KAWs and KSWs,
and investigated properties of low-frequency KAWs, $\omega \ll \omega _{ci}$%
, where $\omega _{ci}$ is the ion cyclotron frequency. Another often made
restricting assumption (see e.g. Shukla \& Stenflo, 2000, and many others)
was that the plasma beta $\beta \ll 1$ ($\beta $ is the plasma/magnetic
pressure ratio). These restrictions make problematic the applicability of
obtained result to the solar wind, where $\beta $ is often $\sim 1$ and the
wave frequency $\omega $ that can approach and exceed $\omega _{ci}$ \cite%
{HU12,SA12}. For the quasi-perpendicular Alfv\'{e}n waves with frequencies
reaching and extending above $\omega /\omega _{ci}=1$, we will still use the
same term KAW. The reason is that the wave dispersion and wave properties do
not change much when the wave frequency crosses $\omega _{ci}$.

In the present study we relax two above restrictions and study KAWs and KSWs
in a wide range of wave and plasma parameters. A two-fluid plasma model is
used to simplify derivations of the wave dispersion and wave properties. The
two-fluid plasma model has been proved to provide a good description for
non-dissipative KAWs \cite{HO99,BE13}. We suggest that the kinetic-scale
PBSs observed in the solar wind can be interpreted not only in terms of
KSWs, but also in terms of KAWs, and hence both these modes can contribute
to PBSs. A more detailed analysis using new mode properties we obtained in
this paper is needed to reveal the dominant mode in every particular event.

In Section 2, we derive the dispersion relation of the waves for the
two-fluid model. Sections 3 and 4 discuss the properties of KAWs and KSWs,
respectively. A discussion and conclusion is given in Section 5. The
detailed derivation of the general dispersion equation is presented in
Appendix A, the wave polarization and correlation properties are given in
Appendix B, and Appendix C gives the analytic expressions of the linear wave
dispersion relations and the linear responses in the low-$\beta $ plasmas.

\section{Dispersion relation and linear response}

We shall start with the linear two-fluid equations
\begin{eqnarray}
m_{\alpha }n_{0}\partial _{t}\mathbf{\delta v}_{\alpha } &=&n_{0}q_{\alpha
}\left( \mathbf{\delta E}+\mathbf{\delta v}_{\alpha }\times \mathbf{B}%
_{0}\right) -\nabla P_{\alpha },  \label{1} \\
\partial _{t}\delta n_{\alpha } &=&-\nabla \cdot \left( n_{0}\mathbf{\delta v%
}_{\alpha }\right) ,  \label{2} \\
\nabla \times \delta \mathbf{B} &=&\mu _{0}\delta \mathbf{J}+\frac{1}{c^{2}}%
\partial _{t}\mathbf{\delta E},  \label{3} \\
\nabla \times \delta \mathbf{E} &=&-\partial _{t}\delta \mathbf{B},
\label{4}
\end{eqnarray}%
where the subscript $\alpha =i,e$ represents ions and electrons,
respectively, $m_{\alpha }$ is the mass, $q_{\alpha }$ is the charge, $%
P_{\alpha }=T_{\alpha }n_{\alpha }$ is the thermal plasma pressure, $%
T_{\alpha }$ is the temperature, $\mathbf{\delta v}_{\alpha }$ is the
perturbed velocity, $\mathbf{B}_{0}=B_{0}\hat{\mathbf{e}}_{z}$ is the
uniform external magnetic field, $\delta \mathbf{J}$ is the perturbed
current density, $\delta \mathbf{E}$ and $\delta \mathbf{B}$ are the
electric and magnetic field perturbations, respectively, $n_{\alpha
}=n_{0}+\delta n_{\alpha }$, $n_{0}$ and $\delta n_{\alpha }$ are the
background and perturbed number densities, respectively. We also assume $%
\omega _{pi}/\omega _{ci}\gg 1$, so that the displacement current in the
Ampere's law (3) can be neglected.

We consider all perturbed variables $\delta f$ in the form of plane waves, $%
\delta f\propto \delta f_{k}\mathbf{exp}(-i\omega t+i\mathbf{k}\cdot \mathbf{%
r})$, where $\omega $ is wave frequency and $\mathbf{k}$ is the wave vector.
The waves are further assumed to propagate in the $(x,z)$ plane, that is, $%
\mathbf{k}=k_{\perp }\hat{\mathbf{e}}_{x}+k_{z}\hat{\mathbf{e}}_{z}$. The
derivation of the general dispersion equation (A17) is given in Appendix A.
Using corresponding approximations, the dispersion equation (A17) can be
reduced to previously derived equations \cite{SH00,ZH10,CH11}. Note that the
two-fluid MHD plasma theory has several limitations as compared to the
kinetic plasma theory. Most importantly, the two-fluid MHD cannot describe
kinetic wave-particle interactions, like Landau damping, transit-time
damping, and cyclotron damping. Also, some wave modes, like ion Bernstein
mode, can only be found in the kinetic theory. As a result, the highly
oblique fast wave transforms into the ion Bernstein mode at $\omega >\omega
_{ci}$\ in the kinetic theory, whereas in the fluid theory it continuously
extends from $\omega <\omega _{ci}$\ to the electron cyclotron frequency $%
\omega \rightarrow \omega _{ce}$\ (Sahraoui et al. 2012). As far as the
highly oblique KAWs\ are concerned, the wave properties given by the
two-fluid MHD are consistent nearly with those given by the kinetic theory
(Sahraoui et al. 2012; Hunana et al. 2013). In particular, the KAW\
dispersion relation is nearly the same in two theories in the low-beta
plasmas (Hunana et al. 2013). The frequency of the quasi-perpendicular slow
mode also rises slowly as compared to the fast mode. Therefore, we focus on
the quasi-perpendicular Alfv\'{e}n and slow modes, but not the fast mode.

For quasi-perpendicular propagation, $k_{\perp }\gg k_{z}$, the cubic
dispersion equation in $\omega ^{2}$ (Equation (A17)) can be reduced to the
quadratic equation for\textbf{\ $\omega ^{2}\ll k^{2}$}$\left(
V_{T}^{2}+V_{A}^{2}\right) $\textbf{:\textbf{\ }}%
\begin{eqnarray}
&&\omega ^{4}\left[ 1+\lambda _{e}^{2}k_{\perp }^{2}+\lambda
_{i}^{2}k_{z}^{2}+\left( 1+\lambda _{e}^{2}k_{\perp }^{2}\right) ^{2}\beta %
\right]   \notag \\
&&-\omega ^{2}\left( 1+2\beta +\rho ^{2}k_{\perp }^{2}\right)
V_{A}^{2}k_{z}^{2}+\beta V_{A}^{4}k_{z}^{4}=0,  \label{5}
\end{eqnarray}%
which describes the dispersion relation of the Alfv\'{e}n and slow waves, $%
\omega ^{2}\sim V_{A}^{2}k_{z}^{2}$ and $\omega ^{2}\sim V_{T}^{2}k_{z}^{2}$%
. Here $\beta =V_{T}^{2}/V_{A}^{2}$. The terms of the order $Q=m_{e}/m_{i}$
or smaller are neglected in Eq. (\ref{5}). The straightforward solutions to
this equation are
\begin{eqnarray}
\omega ^{2} &=&\frac{V_{A}^{2}k_{z}^{2}(1+2\beta +\rho ^{2}k_{\perp }^{2})}{%
2\left( 1+\lambda _{e}^{2}k_{\perp }^{2}+\lambda _{i}^{2}k_{z}^{2}+\left(
1+\lambda _{e}^{2}k_{\perp }^{2}\right) ^{2}\beta \right] }\times   \notag \\
&&\left[ 1\pm \sqrt{1-4\beta \frac{1+\lambda _{e}^{2}k_{\perp }^{2}+\lambda
_{i}^{2}k_{z}^{2}+\left( 1+\lambda _{e}^{2}k_{\perp }^{2}\right) ^{2}\beta }{%
\left( 1+2\beta +\rho ^{2}k_{\perp }^{2}\right) ^{2}}}\right] ,  \label{6}
\end{eqnarray}%
where \textquotedblleft +\textquotedblright\ stands for KAWs and
\textquotedblleft -\textquotedblright\ for KSWs. Dispersion relation derived
by Hollweg (1999) is recovered from (\ref{6}) in the low-frequency limit $%
\lambda _{i}^{2}k_{z}^{2}\ll 1$ (hence $\omega ^{2}\ll \omega _{ci}^{2}$),
and $\left( 1+\lambda _{e}^{2}k_{\perp }^{2}\right) \left( 1+\beta \right)
\simeq 1+\beta +\lambda _{e}^{2}k_{\perp }^{2}$.

\begin{figure}[t]
\epsscale{1.00} \plotone{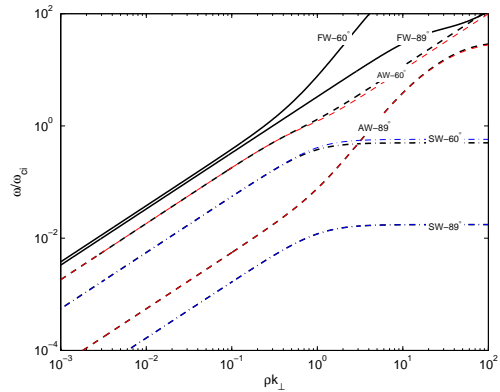}
\caption{Comparison of the dispersion relations (6) (thin dashed lines are
for Alfv\'{e}n waves; thin dot-dashed lines are for slow waves) with the
wave modes obtained from the general dispersion equation (A17) (thick solid
lines are for fast waves; thick dashed lines are for Alfv\'{e}n waves; thick
dot-dashed lines are for slow waves). The two propagation angles are $%
\protect\theta =60^{\circ }$ and $89^{\circ }$, and $\protect\beta =0.1$. }
\label{fig:1}
\end{figure}

Figure 1 compares numerical solutions of the general dispersion equation
(A17) and the dispersion relations (\ref{6}) obtained for the
quasi-perpendicular wave propagation. At such oblique propagation, the fast
mode has significantly higher frequencies than the other two modes. The
remaining KAW and KSW modes behave differently. Similarly to the fast mode,
the KAW mode frequency increases monotonously with increasing wavenumber and
exceeds the $\omega _{ci}$ at some wavelength close to the ion gyroradius.
On the contrary, the KSW frequency slows down it increase above the ion
gyroradius scale and never cross $\omega =\omega _{ci}$. From this figure
one can see that the quasi-perpendicular dispersion relations (\ref{6}) are
also valid down to $\theta \sim 60^{\circ }$.

The physical quantities associated with the KAW and KSW dispersion relations
can be easily obtained from Equations (A1)--(A6) and (A11)--(A15):
\begin{eqnarray}
\delta \mathbf{B} &=&-i\frac{\omega _{ci}}{\omega }\Upsilon _{1}\delta B_{y}%
\hat{\mathbf{e}}_{x}+\delta B_{y}\hat{\mathbf{e}}_{y}+i\frac{k_{\perp }}{%
k_{z}}\frac{\omega _{ci}}{\omega }\Upsilon _{1}\delta B_{y}\hat{\mathbf{e}}%
_{z},  \label{7} \\
\delta \mathbf{E} &=&\frac{\omega }{k_{z}}\left( 1+\lambda
_{e}^{2}k^{2}+\Upsilon _{2}\frac{\widetilde{T}_{e}}{\widetilde{T}_{i}}%
\right) \widetilde{T}_{i}\delta B_{y}\hat{\mathbf{e}}_{x}+i\frac{\omega _{ci}%
}{k_{z}}\Upsilon _{1}\delta B_{y}\hat{\mathbf{e}}_{y}  \notag \\
&&+\frac{\omega }{k_{\perp }}\left( 1+\lambda _{e}^{2}k^{2}+\Upsilon _{2}%
\frac{\widetilde{T}_{e}}{\widetilde{T}_{i}}-\frac{1}{\widetilde{T}_{i}}%
\right) \widetilde{T}_{i}\delta B_{y}\hat{\mathbf{e}}_{z},  \label{8} \\
\delta \mathbf{v}_{i} &=&i\frac{\omega _{ci}}{k_{z}}\left( \Upsilon _{2}-%
\frac{V_{A}^{2}k_{z}^{2}}{\omega ^{2}}\right) \frac{\delta B_{y}}{B_{0}}\hat{%
\mathbf{e}}_{x}-\frac{V_{A}^{2}k_{z}}{\omega }\frac{\delta B_{y}}{B_{0}}\hat{%
\mathbf{e}}_{y}  \notag \\
&&+i\frac{\omega _{ci}}{k_{\perp }}\left( \Upsilon _{2}-1\right) \frac{%
\delta B_{y}}{B_{0}}\hat{\mathbf{e}}_{z},  \label{9} \\
\delta \mathbf{v}_{e} &=&i\frac{\omega _{ci}}{k_{z}}\left( \Upsilon _{2}-%
\frac{V_{A}^{2}k_{z}^{2}}{\omega ^{2}}+\lambda _{i}^{2}k_{z}^{2}\right)
\frac{\delta B_{y}}{B_{0}}\hat{\mathbf{e}}_{x}  \notag \\
&&-\frac{\omega }{k_{z}}\left( 1+\lambda _{e}^{2}k^{2}-Q\frac{%
V_{A}^{2}k_{z}^{2}}{\omega ^{2}}\right) \frac{\delta B_{y}}{B_{0}}\hat{%
\mathbf{e}}_{y}  \notag \\
&&+i\frac{\omega _{ci}}{k_{\perp }}\left( \Upsilon _{2}-\lambda
_{i}^{2}k_{\perp }^{2}-1\right) \frac{\delta B_{y}}{B_{0}}\hat{\mathbf{e}}%
_{z},  \label{10}
\end{eqnarray}%
where
\begin{eqnarray}
\Upsilon _{1} &=&\left( 1+\lambda _{e}^{2}k^{2}\right) \frac{\omega ^{2}}{%
V_{A}^{2}k^{2}}-\frac{k_{z}^{2}}{k^{2}},  \notag \\
\Upsilon _{2} &=&\left( 1+\lambda _{e}^{2}k^{2}\right) ^{2}\frac{\omega ^{2}%
}{V_{A}^{2}k^{2}}-\frac{k_{z}^{2}}{k^{2}}+\frac{V_{A}^{2}k_{z}^{2}}{\omega
^{2}}-\lambda _{i}^{2}k_{z}^{2},  \notag
\end{eqnarray}%
The number density and the parallel magnetic field are related by
\begin{equation}
\frac{\delta n}{n_{0}}=-\frac{1}{\beta }\frac{k^{2}}{k_{\perp }^{2}}\frac{%
1+\lambda _{e}^{2}k^{2}-\Upsilon _{2}}{1+\lambda
_{e}^{2}k^{2}-V_{A}^{2}k_{z}^{2}/{\omega ^{2}}}\frac{\delta B_{z}}{B_{0}}.
\end{equation}%
These relations (7)--(11) can be used in the diagnostics of experimentally
observed wave phenomena.

\section{KAW properties}

KAWs behave differently in different $\beta $ regimes, namely, the inertial
regime $\beta <m_{e}/m_{i}$, the kinetic regime $m_{e}/m_{i}<\beta <1$, and
the high-$\beta $ regime $\beta \gtrsim 1$. Thus, we will investigate the
KAW properties for the representative values $\beta =10^{-4}$, typical in
the Earth's ionosphere and the solar flare loops, $\beta =10^{-2}$, typical
in the Earth's magnetosphere and the solar corona, and $\beta =1$, typical
in the solar wind at $\sim 1$AU.

\begin{figure*}[t]
\epsscale{1.00} \plotone{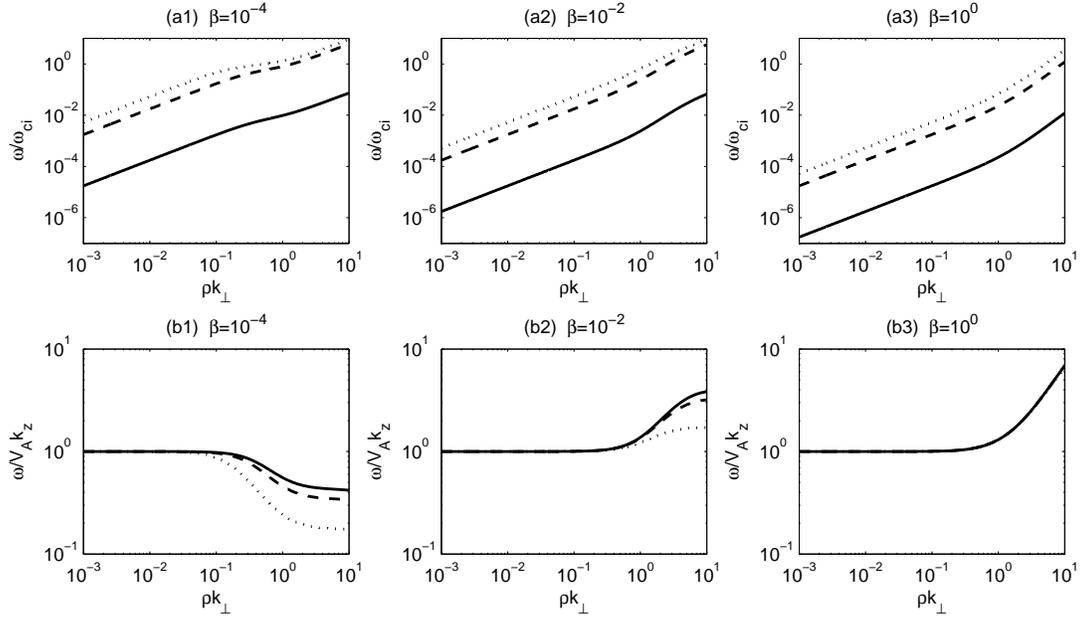}
\caption{Wave frequency of KAWs in three different $\protect\beta $ regimes:
inertial regime $\protect\beta =10^{-4}$, kinetic regime $\protect\beta %
=10^{-2}$, and high-$\protect\beta $ regime $\protect\beta =1$. The dotted,
dashed, and solid lines represent the propagating angle $\protect\theta %
=87^{\circ },89^{\circ }$, and $89^{\circ }.99$, respectively. (a) $\protect%
\omega $ is normalized by $\protect\omega _{ci}$; (b) $\protect\omega $
normalized by $V_{A}k_{z}$. The equal electron and ion temperatures, $%
T_{e}=T_{i}$, is used here and following figures.}
\label{fig2}
\end{figure*}

From Figure 2 one can see that the\textbf{\ }KAW frequency is larger than
the ion-cyclotron frequency, $\omega \gtrsim \omega _{ci}$, when $k_{\perp
}\rho =1$\ and $\theta =87^{\circ }$ or $89^{\circ }$, but $\omega /\omega
_{ci}<1$ for extremely oblique propagation, $\theta =89^{\circ }.99$. This
is consistent with the result in Sahraoui et al. (2012) that $\omega /\omega
_{ci}<1$ at all scales as the propagating angle $\theta >\theta _{\mathrm{%
crit}}=\mathrm{cos}^{-1}(Q)\simeq 89^{\circ }.97$.

\begin{figure*}[t]
\epsscale{1.00} \plotone{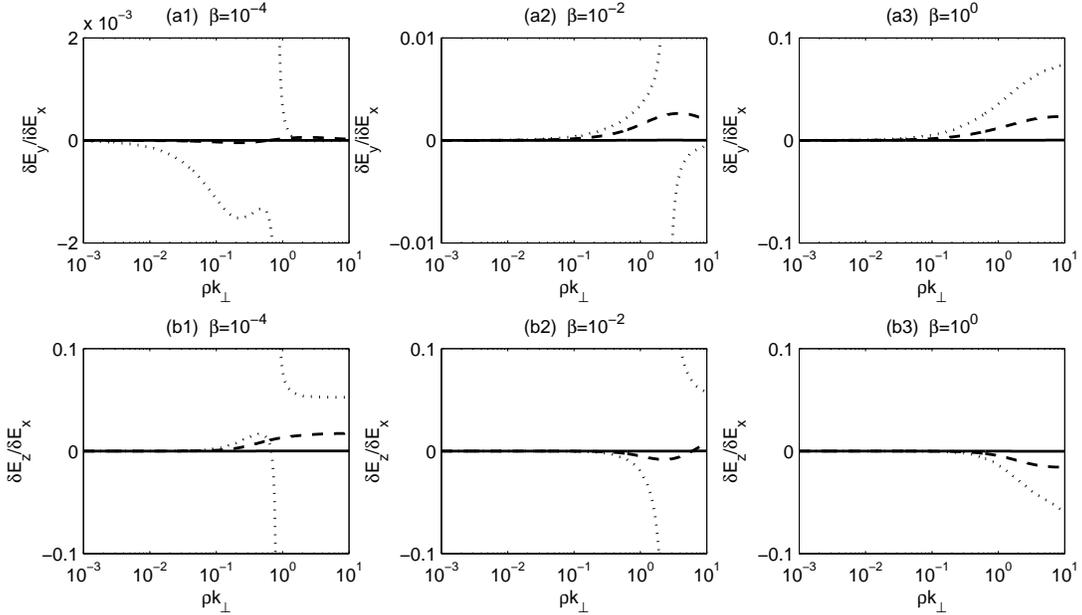}
\caption{Electric polarization ratios $\protect\delta E_{y}/(i\protect\delta %
E_{x})$ and $\protect\delta E_{z}/\protect\delta E_{x}$ for KAWs. The
dotted, dashed, and solid lines represent the propagating angles $\protect%
\theta =87^{\circ },$ $89^{\circ }$, and $89^{\circ }.99$, respectively.}
\label{fig:3}
\end{figure*}

Figure 3 presents the electric polarization of KAWs. The waves are polarized
elliptically (almost linearly in low-$\beta $ plasmas), $P_{\delta E,\mathbf{%
B}_{0}}=\delta E_{y}/\left( i\delta E_{x}\right) <1$. At relatively small $%
\rho k_{\perp }$,\ the polarization parameter is positive, $P_{\delta E,%
\mathbf{B}_{0}}>0$, in the kinetic and high-$\beta $ regimes, which
corresponds to the right-hand polarization, as was first shown by \cite%
{GA86,HO99}. However, at larger $\rho k_{\perp }>1$\ there are several
transition points where $\delta E_{x}$ passes through zero and $\delta
E_{y}/i\delta E_{x}$ and $\delta E_{z}/\delta E_{x}$ change their signs.
These polarization reversals are discussed in more detail below.

\begin{figure*}[h]
\epsscale{1.00} \plotone{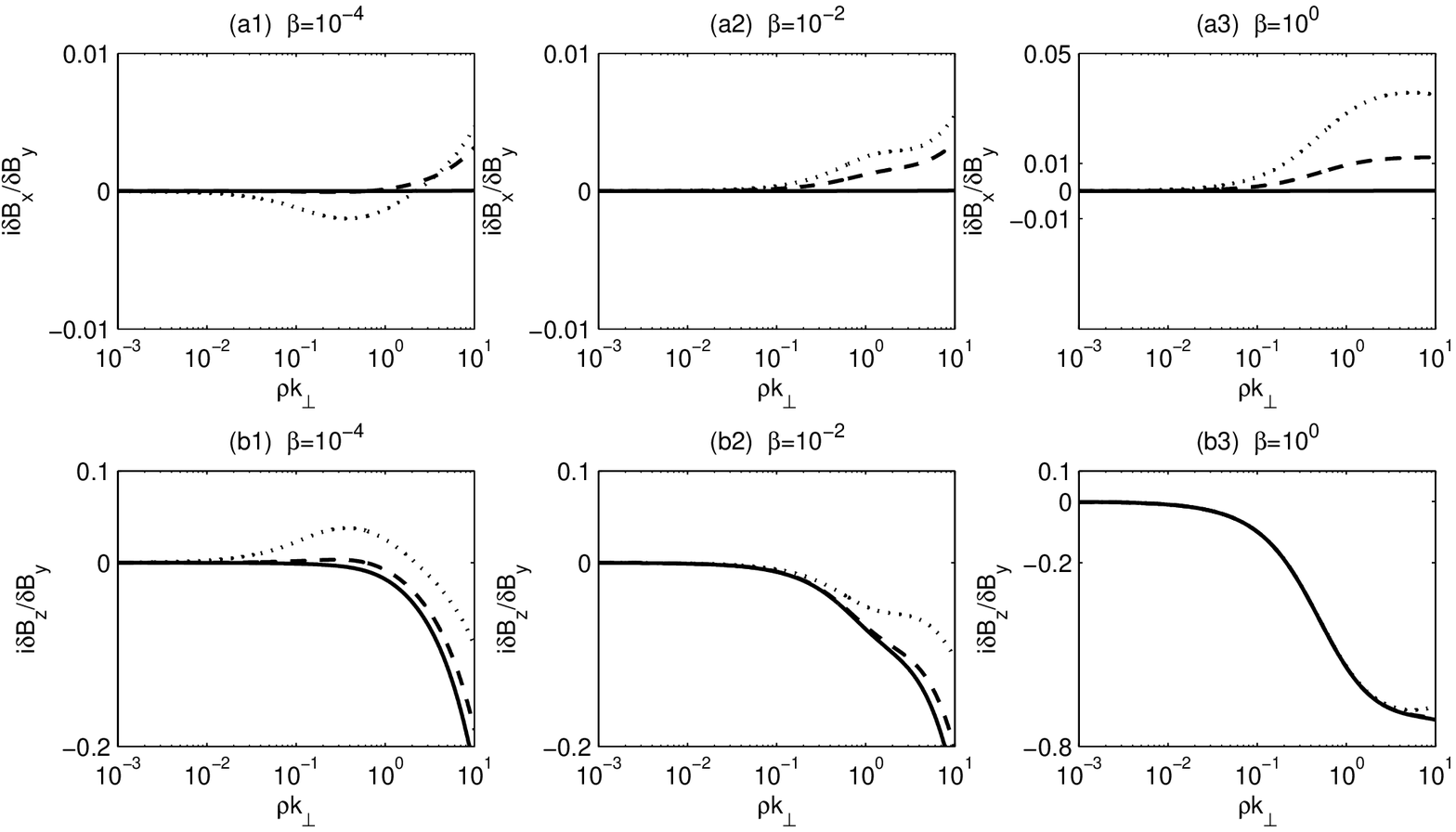}
\caption{Magnetic polarization ratios $i\protect\delta B_{x}/\protect\delta %
B_{y}$ and $i\protect\delta B_{z}/\protect\delta B_{y}$ for KAWs. The
dotted, dashed, and solid lines represent the wave propagation angles $%
\protect\theta =87^{\circ },$ $89^{\circ }$, and $89^{\circ }.99$,
respectively.}
\label{fig:4}
\end{figure*}

Figures 4 and 5 show the magnetic polarization and the magnetic helicity $%
\sigma $ of KAWs. At the ion scale $\rho k_{\perp }\sim 1$ we observe quite
small $i\delta B_{x}/\delta B_{y}\sim 0.01$, but the values of $i\delta
B_{z}/\delta B_{y}$ are larger. For $\rho k_{\perp }\lesssim 1$ the KAW
helicity is right-hand $(\sigma <0)$ in the inertial regime but becomes
left-hand at larger $\beta $\ (it also becomes left-hand in the inertial
range at larger $\rho k_{\perp }$).\textbf{\ }

\begin{figure*}[h]
\epsscale{1.00} \plotone{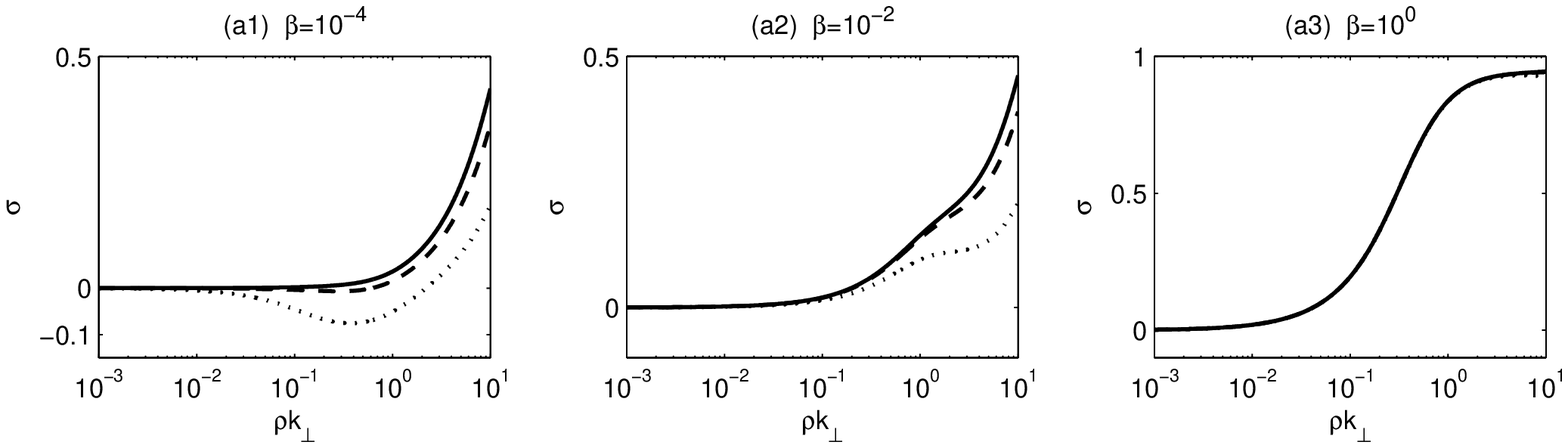}
\caption{Magnetic helicity $\protect\sigma $ of KAWs. The dotted, dashed,
and solid lines represent the propagation angles $\protect\theta =87^{\circ
},$ $89^{\circ }$, and $89^{\circ }.99$, respectively.}
\label{fig:5}
\end{figure*}

\begin{figure*}[h]
\epsscale{1.00} \plotone{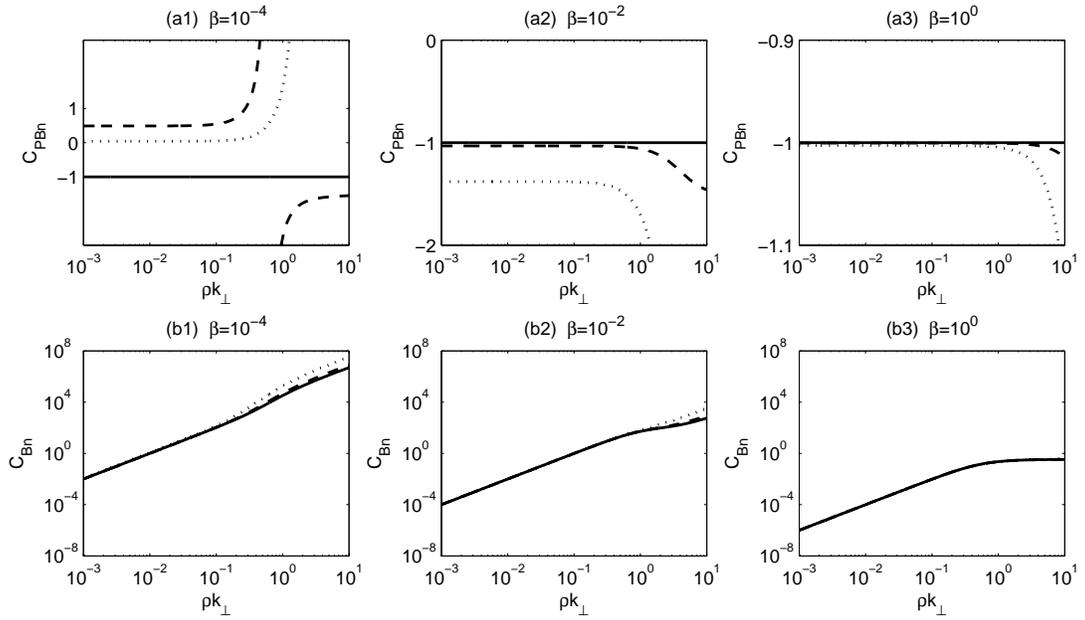}
\caption{Pressure correlation $C_{PBn}$ and compressibility $C_{Bn}$ of
KAWs. The dotted, dashed, and solid lines represent the propagation angles $%
\protect\theta =87^{\circ },$ $89^{\circ }$, and $89^{\circ }.99$,
respectively.}
\label{fig:6}
\end{figure*}

\begin{figure*}[t]
\epsscale{1.00} \plotone{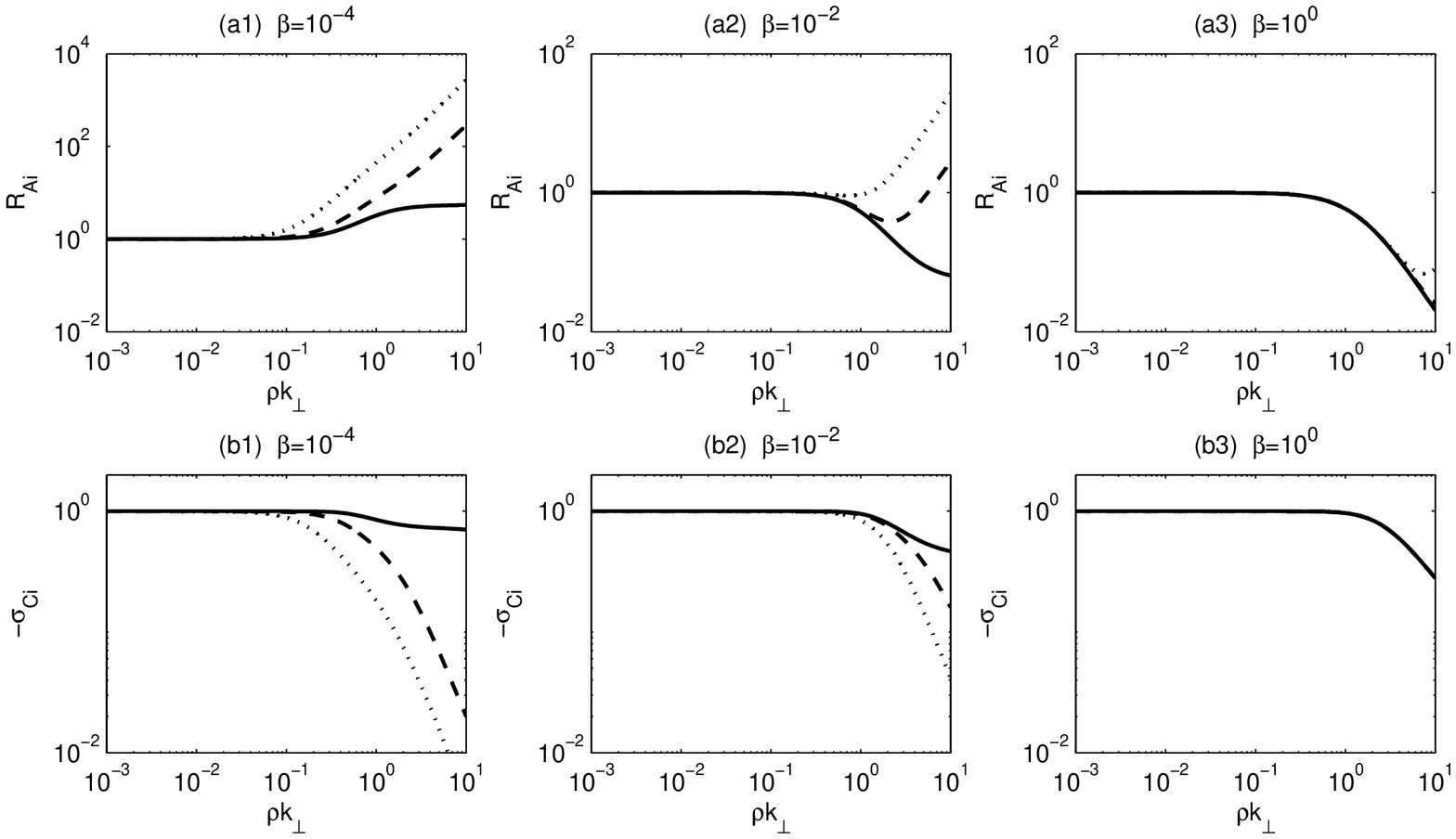}
\caption{Ion Alfv\'{e}n ratio $R_{Ai}$ and cross helicity $\protect\sigma %
_{Ci}$ of KAWs. The dotted, dashed, and solid lines represent the
propagation angles $\protect\theta =87^{\circ },89^{\circ }$, $89^{\circ }$,
and $89^{\circ }.99$, respectively.}
\label{fig:7}
\end{figure*}

Figure 6 presents the plasma-magnetic pressure correlation $C_{PBn}=\frac{%
\gamma \left( T_{i}+T_{e}\right) \delta n}{\delta B_{z}B_{0}/\mu _{0}}$ and
compressibility $C_{Bn}=\frac{\delta n^{2}/n_{0}^{2}}{\delta B^{2}/B_{0}^{2}}
$ of KAWs. It is interesting to observe the pressure balance $C_{PBn}\simeq
-1$ in the extremely oblique KAWs $(\theta =89^{\circ }.99)$ at arbitrary $%
\beta $. $C_{PBn}\simeq -1$ also holds for the arbitrary propagation angle
in the high-$\beta $ plasmas. Note several transition points in the inertial
regime where $C_{Bn}$ changes its sign, which occur when $k_{z}^{2}/k_{\perp
}^{2}-\left( 1+\lambda _{e}^{2}k_{\perp }^{2}\right) \beta =0$. $C_{Bn}$
nearly follows the approximate expression $C_{Bn}\simeq \rho ^{2}k_{\perp
}^{2}\left( 1+\lambda _{e}^{2}k_{\perp }^{2}+\lambda
_{i}^{2}k_{z}^{2}\right) /\beta \left( {1+\rho ^{2}k_{\perp }^{2}}\right) $
that is valid for the low $\beta $.

Figure 7 presents the ion Alfv\'{e}n ratio $R_{Ai}$ and the ion cross
helicity $\sigma _{Ci}$ of KAWs. $R_{Ai}$ and $\sigma _{Ci}$ can be
rewritten as $R_{Ai}=\delta v_{i}^{2}/\delta v_{B}^{2}$ and $\sigma
_{Ci}=2\left( \delta \mathbf{v}_{i}\cdot \delta \mathbf{v}_{B}^{\ast
}\right) /\left( \delta v_{i}^{2}+\delta v_{B}^{2}\right) $, where $\delta
v_{B}$ is the magnetic perturbation in the units of Alfv\'{e}n speed. The
ion cross-helicity becomes nearly zero, $\sigma _{Ci}\sim 0$, as the strong
velocity mismatch appears, $\delta v_{i}\gg \delta v_{B}$ (for the kinetic
scale $\lambda _{e}k_{\perp }>1$ waves with $\rho k_{\perp }>10^{-2}$ in the
inertial regime) or $\delta v_{B}\gg \delta v_{i}$ (for KAWs with $\rho
k_{\perp }>1$ in the high-$\beta $ plasmas).

In the following sub-sections we consider the electric field polarization
and its reversal in more detail.

\subsection{Electric Polarization and Its Reversal}

In the limit $k_{\perp }\gg k_{z}$, from the expression (\ref{8}) we get the
polarization ratio
\begin{equation}
\frac{\delta E_{y}}{\delta E_{x}}=i\frac{\omega }{\omega _{ci}}\frac{1}{%
\lambda _{i}^{2}k_{\perp }^{2}}\frac{\left( 1+\lambda _{e}^{2}k_{\perp
}^{2}\right) \frac{\omega ^{2}}{V_{A}^{2}k_{z}^{2}}-1}{\widetilde{T}%
_{i}\left( 1+\lambda _{e}^{2}k_{\perp }^{2}\right) \frac{\omega ^{2}}{%
V_{A}^{2}k_{z}^{2}}+\widetilde{T}_{e}\left( 1-\frac{\omega ^{2}}{\omega
_{ci}^{2}}\right) }.  \label{pol1}
\end{equation}%
The sense of the wave polarization can change when the numerator or
denominator of this expression passes through zero, which correspond to $%
\delta E_{y}=0$ or $\delta E_{x}=0$ , respectively. The $\delta E_{x}=0$
transition occurs at
\begin{equation}
\left( \frac{\omega ^{2}}{\omega _{ci}^{2}}-1\right) =\frac{\widetilde{T}_{i}%
}{\widetilde{T}_{e}}\left( 1+\lambda _{e}^{2}k_{\perp }^{2}\right) \frac{%
\omega ^{2}}{V_{A}^{2}k_{z}^{2}}>0,  \label{x.tr}
\end{equation}%
which implies high-frequency $\omega >\omega _{ci}$ waves at the transition
point. For the low-frequency waves ($\omega \ll \omega _{ci}$), there are no
such transition points. In low-$\beta $ limit this transition can occur as
\begin{equation}
\lambda _{i}^{2}k_{z}^{2}=\frac{\left( 1+\lambda _{e}^{2}k_{\perp
}^{2}\right) \left( 1+\left( 1+\rho ^{2}k_{\perp }^{2}\right) \widetilde{T}%
_{i}/\widetilde{T}_{e}\right) }{\rho ^{2}k_{\perp }^{2}}.  \label{theta1}
\end{equation}

The $\delta E_{y}=0$ transition occurs at
\begin{equation}
\frac{\omega ^{2}}{V_{A}^{2}k_{z}^{2}}=\frac{1}{1+\lambda _{e}^{2}k_{\perp
}^{2}}<1.  \label{y.tr}
\end{equation}%
This transition implies sub-Alfv\'{e}nic phase velocities of KAWs and is
possible only due to finite $\lambda _{i}^{2}k_{z}^{2}$. In low-$\beta $
limit (\ref{y.tr}) gives the transition wavenumber
\begin{equation}
\rho ^{2}k_{\perp \mathrm{tr}}^{2}=\frac{m_{i}}{m_{e}}\left( \frac{1}{\tan
^{2}\theta }-\beta \right) .
\end{equation}%
With growing $\rho ^{2}k_{\perp }^{2}$, the transition occurs from the left-
to right-hand polarization. For this transition to occur, the wave
propagation angle should be less than certain value,
\begin{equation}
\theta <\theta _{\mathrm{cr}}=\arctan \frac{1}{\sqrt{\beta }}.
\label{theta2}
\end{equation}%
Otherwise, the wave is always right-hand polarization. Only in this last
case the conclusion by \cite{GA86,HO99} holds that KAWs are right-hand
polarized.

The above analysis indicates that KAWs can be both left- and right-hand
polarized.

\subsection{The Low-$\protect\beta $ Low-frequency Limit}

For KAWs in low-$\beta $ plasmas, $\beta \ll 1$, we get
\begin{equation}
\frac{\delta E_{y}}{\delta E_{x}}=i\beta \frac{\omega }{\omega _{ci}}\frac{1%
}{1+\rho _{i}^{2}k_{\perp }^{2}}\frac{\left( 1+\lambda _{e}^{2}k_{\perp
}^{2}\right) -\lambda _{i}^{2}k_{z}^{2}/\rho ^{2}k_{\perp }^{2}}{\left(
1+\lambda _{e}^{2}k_{\perp }^{2}\right) -\lambda _{i}^{2}k_{z}^{2}\rho
_{s}^{2}k_{\perp }^{2}/\left( 1+\rho _{i}^{2}k_{\perp }^{2}\right) }.
\label{pol2}
\end{equation}%
At small $\lambda _{i}^{2}k_{z}^{2}\ll 1$ (i.e. in the low-frequency range)
the denominator of the above expression is dominated by the term $\left(
1+\lambda _{e}^{2}k_{\perp }^{2}\right) $, and we arrive to
\begin{equation}
\frac{\delta E_{y}}{\delta E_{x}}=i\beta \frac{\omega }{\omega _{ci}}\frac{1%
}{1+\rho _{i}^{2}k_{\perp }^{2}}\left( 1-\frac{1}{\left( \beta +\frac{m_{e}}{%
m_{i}}\rho ^{2}k_{\perp }^{2}\right) \tan ^{2}\theta }\right) ,  \label{pol3}
\end{equation}%
which indicates that the electric polarization depends on the magnitude of $%
k_{\perp }$ compared to the ion kinetic scales. In the limit $\beta \tan
^{2}\theta \gg 1$, our expression (\ref{pol3}) simplifies to
\begin{equation}
\frac{\delta E_{y}}{\delta E_{x}}=i\beta \frac{\omega }{\omega _{ci}}\frac{1%
}{1+\rho _{i}^{2}k_{\perp }^{2}}.  \label{pol4}
\end{equation}%
In this limit the waves are always right-hand polarized. In principle, this
conclusion agrees with previous results \cite{GA86,HO99}. In the
long-wavelength limit $\rho _{i}^{2}k_{\perp }^{2}\ll 1$, Equation (\ref%
{pol3}) reduces to the approximate analytical formula Equation (46) by \cite%
{HO99},
\begin{equation}
\frac{\delta E_{x}}{\delta E_{y}}=i\beta \frac{\omega }{\omega _{ci}}.
\label{pol.hol}
\end{equation}

\section{KSW properties}

Before discussing properties of the oblique slow waves in the two-fluid MHD,
it is instructive to mention some their known properties in the kinetic
theory. At general oblique propagation, the slow/sound wave extends to the
frequency larger than the ion cyclotron frequency as showed in Figure 1 by
Krauss-Varban et al. (1994) for the propagation angle $\theta =30^{\circ }$.
At larger propagation angles, large wavenumbers are required for the waves
to reach the ion-cyclotron frequency, and for quasi-perpendicular
propagation slow waves remain sub-cyclotron in the wide range of
perpendicular wavenumber, up to large values of $\rho ^{2}k_{\perp }^{2}$.
In the two-fluid MHD, the frequency of quasi-perpendicular KSWs always
remains sub-cyclotron:\textbf{\ }%
\begin{equation*}
\frac{\omega }{\omega _{ci}}=\frac{k_{z}}{k_{\perp }}\sqrt{\frac{\rho
^{2}k_{\perp }^{2}}{1+\rho ^{2}k_{\perp }^{2}}}<1\mathrm{\quad for\quad }%
\frac{k_{z}}{k_{\perp }}\leq 1.
\end{equation*}

\begin{figure*}[h]
\epsscale{0.90} \plotone{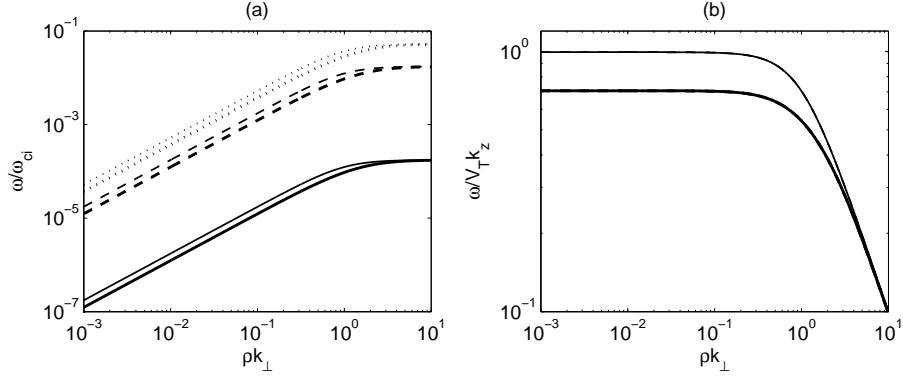}
\caption{Wave frequency of KSWs in three different $\protect\beta $ regimes,
the inertial regime ($\protect\beta =10^{-4}$, the thin lines), the kinetic
regime ($\protect\beta =10^{-2}$, the middle lines), and the high-$\protect%
\beta $ regime ($\protect\beta =1$, the thick lines), where the dotted,
dashed, and solid lines represent the propagating angles $\protect\theta %
=87^{\circ },$ $89^{\circ }$, and $89^{\circ }.99$, respectively. The lines
in the inertial regime are the same as in the kinetic regime. (a) $\protect%
\omega $ is normalized by $\protect\omega _{ci}$; (b) $\protect\omega $
normalized by $V_{T}k_{z}$.}
\label{fig:8}
\end{figure*}

The KSW dispersion is showed in Figure 8 in terms of the phase velocity $%
\omega /\left( V_{T}k_{z}\right) $, which exhibits also a depression at high
$\beta $\ in the long-wavelength limit.\textbf{\ }

\begin{figure*}[tbp]
\epsscale{0.90} \plotone{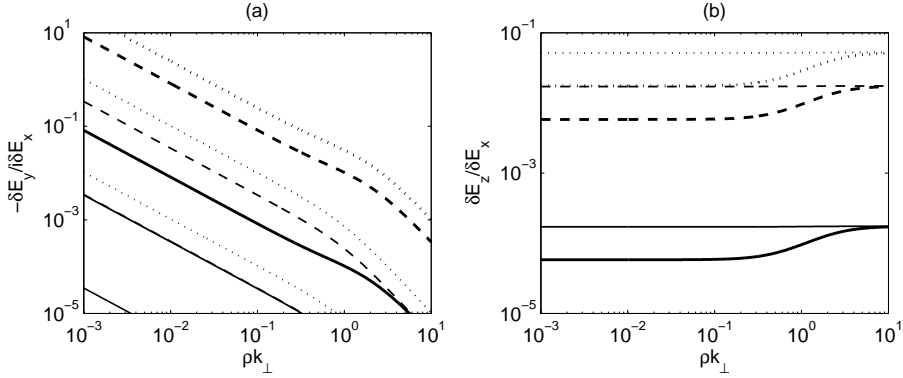}
\caption{Electric-field polarization $\protect\delta E_{y}/\left( i\protect%
\delta E_{x}\right) $ and $\protect\delta E_{z}/\protect\delta E_{x}$ of
KSWs, where the lines have the same meaning as in Figure 8.}
\label{fig:9}
\end{figure*}

Figure 9 presents the electric polarization of KSWs. The polarization
parameter $P_{E,B_{0}}=E_{y}/\left( i\delta E_{x}\right) <0$ means the
left-hand KSW polarization. $\delta E_{x}$ becomes the dominant component
for the waves at the ion gyroradius scale $(\rho k_{\perp }\sim 1)$. The KSW
electric field polarization ratios can be approximated as $i\delta
E_{y}/\delta E_{x}\simeq \beta k_{z}/\left( {\rho k_{\perp }^{2}\sqrt{1+\rho
^{2}k_{\perp }^{2}}}\right) $, and $\delta E_{z}/\delta E_{x}=k_{z}/k_{\perp
}$.

\begin{figure*}[tbp]
\epsscale{0.90} \plotone{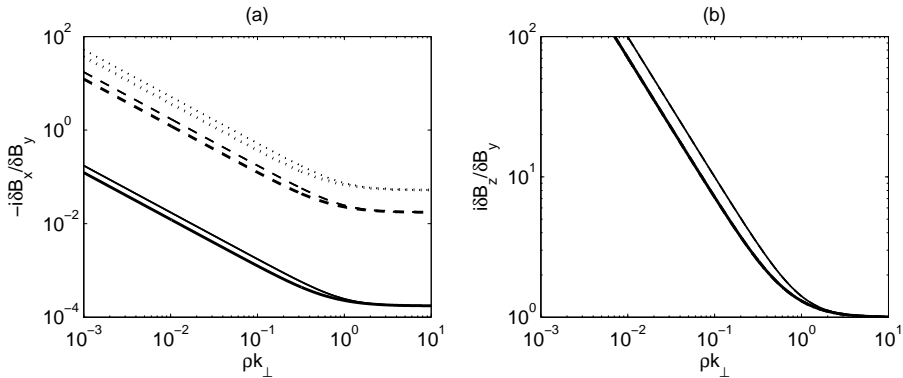}
\caption{Magnetic-field polarization $i\protect\delta B_{x}/\protect\delta %
B_{y}$ and $\protect\delta B_{z}/\protect\delta B_{y}$ of KSWs, where the
lines have the same meaning as in Figure 8.}
\label{fig:10}
\end{figure*}

Figure 10 presents the magnetic polarization of KSWs. For the long
wavelength waves ($\rho k_{\perp }\ll 1$), the compressible magnetic field
perturbation $\delta B_{z}$ is the dominant component, $|\delta B_{z}/\delta
B_{x,y}|\gg 1$. However, $\delta B_{y}$ becomes as important as $\delta
B_{z} $ when the wavelength approaches the ion gyroradius scale, where $%
\delta B_{y}\simeq \delta B_{z}\gg \delta B_{x}$. The magnetic field
polarization ratios nearly follow the approximate relations $i\delta
B_{x}/\delta B_{y}=-\left( k_{z}/k_{\perp }\right) \sqrt{1+1/\rho
^{2}k_{\perp }^{2}}$ and $i\delta B_{z}/\delta B_{y}=\sqrt{1+1/\rho
^{2}k_{\perp }^{2}}$.

The helicity $\sigma $ in the low-$\beta $ plasmas can be written as $\sigma
=-2\rho k_{\perp }\left( 1+\rho ^{2}k_{\perp }^{2}\right) ^{1/2}/\left(
1+2\rho ^{2}k_{\perp }^{2}\right) $, which indicates that $\sigma $
decreases from $0$ to $-1$ as $\rho k_{\perp }$ increases from $10^{-3}$ to $%
10$. This behavior is seen from Figure 11.

\begin{figure}[tbp]
\epsscale{1.00} \plotone{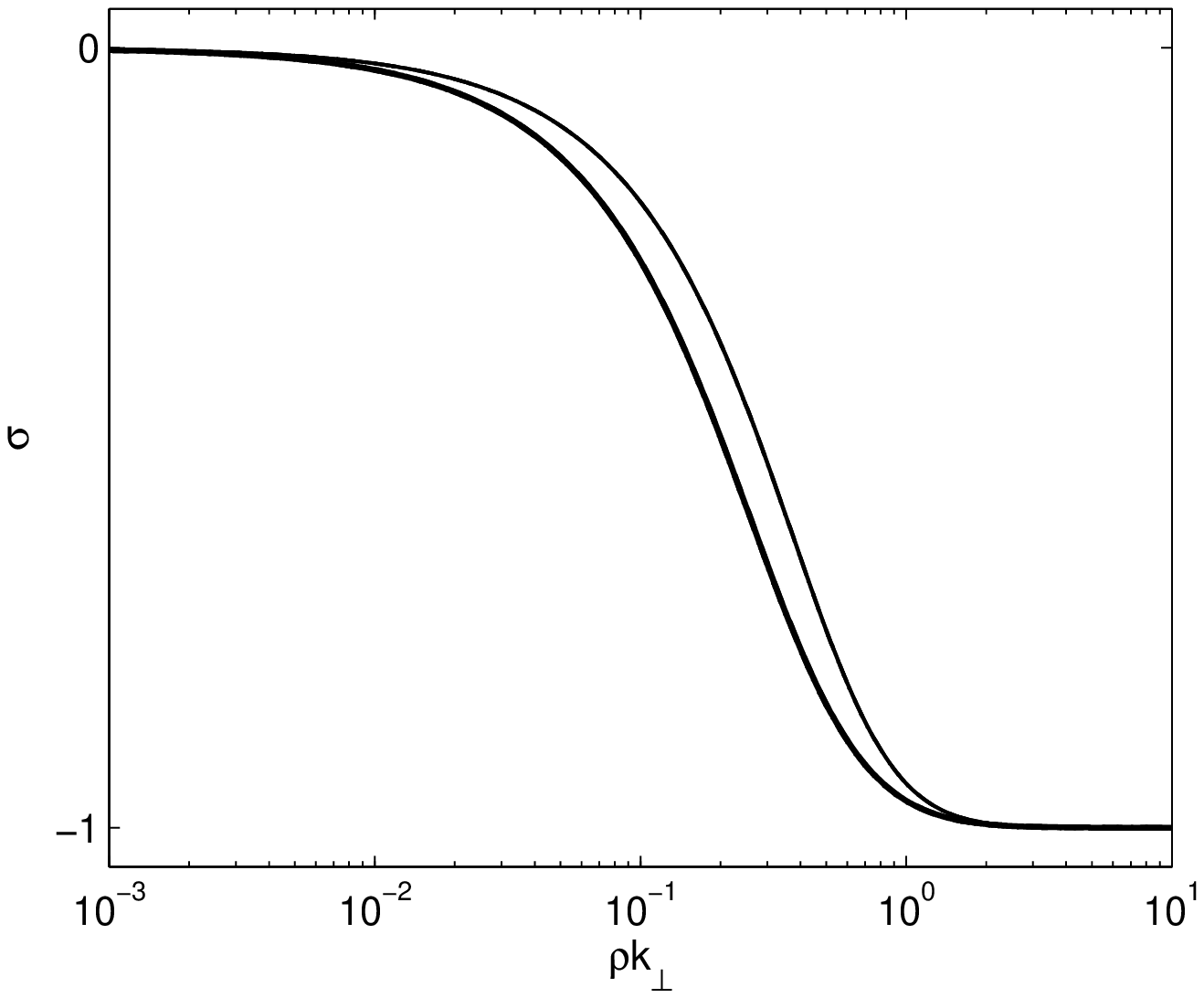}
\caption{Magnetic helicity of KSWs. The lines have the same meaning as in
Figure 8.}
\label{fig:11}
\end{figure}

Figure 12 presents the plasma/magnetic pressure correlation $C_{PBn}$ and
the compressibility $C_{Bn}$ of KSWs. The behavior of these functions is in
accordance with the theoretic predictions that $C_{PBn}\simeq -1$ and $%
C_{Bn}\simeq \left( 1+\rho ^{2}k_{\perp }^{2}\right) /\left( 1+2\rho
^{2}k_{\perp }^{2}\right) /\beta ^{2}$ in the low-$\beta $ plasma. Note that
$C_{PBn}$ and $C_{Bn}$ in the high-$\beta $ plasmas can be approximated by
the same expressions.

\begin{figure*}[t]
\epsscale{0.90} \plotone{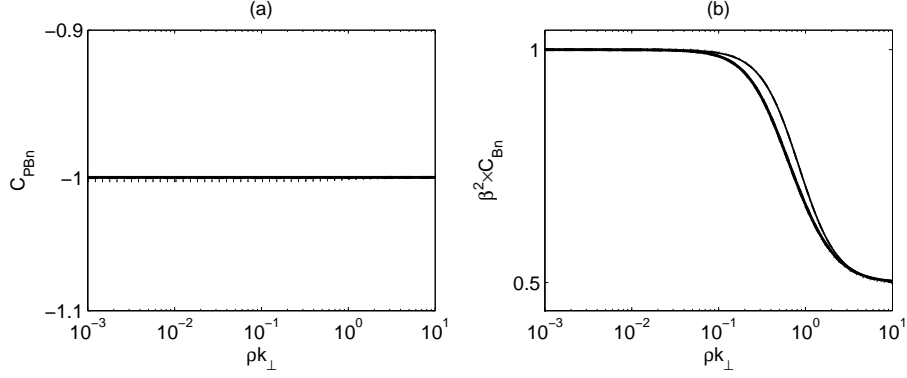}
\caption{Plasma/magnetic pressure ratio $C_{PBn}$ and compressibility $%
C_{Bn} $ of KSWs. The lines have the same meaning as in Figure 8.}
\label{fig:12}
\end{figure*}

Figure 13 presents the ion Alfv\'{e}n ratio $R_{Ai}$ and the cross-helicity $%
\sigma _{Ci}$ of KSWs. $\delta v_{i}$ is nearly equal $\delta v_{B}$ in the
long-wavelength waves $(\rho k_{\perp }\ll 1)$ in high-$\beta $ plasmas. In
other cases $\delta v_{i}$ dominates over $\delta v_{B}$, $\delta
v_{i}>\delta v_{B}$. The corresponding expressions in the low-$\beta $
plasmas are $R_{Ai}\simeq \left( 1+2\rho ^{2}k_{\perp }^{2}\right) /\beta $
and $\sigma _{Ci}\simeq -2\sqrt{\beta /\left( 1+\rho ^{2}k_{\perp
}^{2}\right) }$.

\begin{figure*}[t]
\epsscale{0.90} \plotone{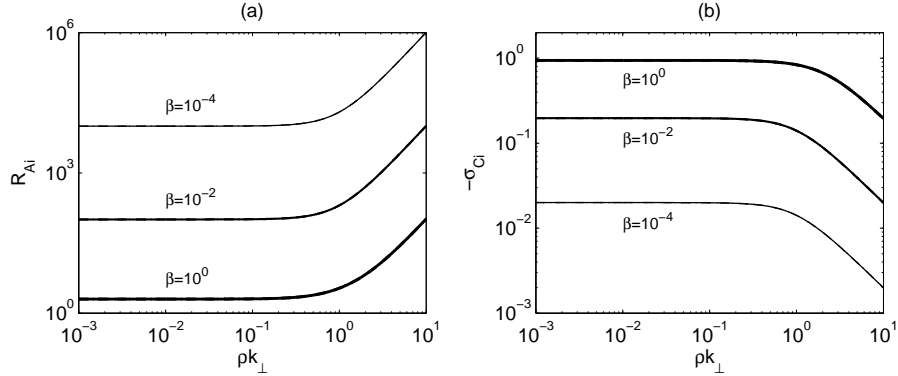}
\caption{Ion Alfv\'{e}n ratio $R_{Ai}$ and cross-helicity $\protect\sigma %
_{Ci}$ of KSWs. The lines have the same meaning as in Figure 8.}
\label{fig:13}
\end{figure*}

\section{Discussion and conclusion}

Our study shows that the Alfv\'{e}n wave frequency $\omega $ at the ion
gyroscale $\rho k_{\perp }\sim 1$ is smaller than the ion-cyclotron
frequency $\omega _{ci}$ for extremely oblique propagation, say for $\theta
=89^{\circ }.99$. In agreement with (Sahraoui et al. 2012), the KAW
frequency can reach $\omega _{ci}$ for propagation angle $\theta \leq
89^{\circ }.97$. At $\rho k_{\perp }\sim 1$, $\omega $ can reach and exceed $%
\omega _{ci}$ for the propagation angles $\sim 87^{\circ }$ or less. At
smaller propagating angles, the frequency $\omega \sim \omega _{ci}$ occurs
at smaller wavenumbers. In the solar-terrestrial plasmas, the high-frequency
KAWs may be generated through the Alfv\'{e}nic turbulent cascade \cite{HU12}%
, excited kinetically by the field-aligned currents and ion beams \cite%
{VO02,VO03}, or by phase mixing combined with cyclotron sweep of Alfv\'{e}n
waves \cite{VO06} .

Our study investigated KAWs and KSWs in a wide range of $\beta ,$ covering a
wide range of conditions in most solar-terrestrial plasmas. Several mode
properties of KAWs are totally different in the limits $\beta <m_{e}/m_{i}$
(so-called inertial range) and $\beta >m_{e}/m_{i}$ (kinetic range). So, we
confirmed that the KAW phase velocity $\omega /k_{z}$ in the inertial range
decreases with increasing $k_{\perp }$, but increases in the kinetic and
high-$\beta $ ranges \cite{GO79,LY96,ST00}). The electric polarization
ratios of KAWs are also obviously different in these two distinct $\beta $
ranges. Namely, except for the case of extremely oblique waves, at smaller $%
k_{\perp }$ KAWs are left-hand polarized in the inertial range and
right-hand in the kinetic/high-$\beta $ ranges. The magnetic helicity at
these wavenumbers is right-hand for $\beta <m_{e}/m_{i}$ and left-hand for $%
\beta >m_{e}/m_{i}$. At higher $k_{\perp }$ they undergo two polarization
reversals defined by (\ref{x.tr}) and (\ref{y.tr}). These new polarization
properties of KAWs, especially polarization reversals, are quite specific
and can be used as a critical test for the mode identification in the solar
wind and terrestrial magnetosphere. In addition to the Landau dissipation
caused by the parallel electric field fluctuations of KAWs, the
perpendicular electric-field can induce the occurrence of the cyclotron
resonant damping as the frequency of KAWs\ reaches or exceeds the ion
cyclotron frequency (Voitenko \& Goossens 2002, 2003). The
cyclotron-resonant damping (Kennel \& Wong 1967; Marsch 2006) does not
require purely left-hand or purely right-hand polarizations of Alfv\'{e}n
waves. When the resonant-cyclotron condition $\omega \left( k_{z}\right)
-k_{z}v_{z}=n\omega _{ci}$\ is satisfied for the oblique waves, there are
two resonance cases depending on the interaction with $E_{x}$\ or $E_{y}$\
(Hollweg \& Markovskii 2002), where $v_{z}$\ is the parallel velocity of
particles. $E_{x}$\ can cause the cyclotron resonance if the particles stay
in the phase with the waves, whereas the resonance relating to $E_{y}$\
strongly depends on the x-position of the particles (Hollweg \& Markovskii
2002). Both Landau and cyclotron dampings are crucial when investigating the
turbulence dissipation channels at kinetic scales. On the contrary, the KSWs
properties are nearly the same in all $\beta $ ranges and can be
approximately described by the approximate expressions for the low-$\beta $
plasmas given in Section 4.

Our study provided a clear evidence of anti-correlation between the plasma
and magnetic pressures for both KAWs and KSWs in high-$\beta $ plasmas. This
makes the total pressure fluctuations almost zero, $\delta P_{tot}\simeq 0$,
which resembles the main common property of the observed PBSs \cite{YAO11}.
Kellogg \& Horbury (2005) and Yao et al. (2011) used the Cluster data, which
have a high time resolution of 0.2 s for the plasma number density and
magnetic field, and found that the scales of PBSs extend down to the ion
scale. Kellogg \& Horbury (2005) interpreted these kinetic-scale PBSs in
terms of KSWs. However, KAWs may be an alternative explanation since they
also drive anti-correlated number density and magnetic field fluctuations.
Again, one needs more mode properties to discriminate which mode dominates
in PBSs, KSW or KAW. Our results reveal three different properties that
distinguish these modes at the ion gyroradius scale: (1) right-hand electric
polarization for KAWs and left-hand for KSWs; (2) magnetic helicity $\sigma
\sim 1$ for KAWs and $\sigma \sim -1$ for KSWs; (3) Alfv\'{e}n ratio $%
R_{Ai}\ll 1$ for KAWs and $R_{Ai}\gg 1$ for KSWs. Besides, the large-scale
Alfv\'{e}n waves, permeating the solar wind, can nonlinearly excite
simultaneous KSWs and KAWs \cite{ZH14}, which both can contribute to the
observed PBSs. This implies that PBSs may be KSWs, or KAWs, or a mixture of
KSWs and KAWs. A recent work (Hollweg et al. 2014) showed that the
highly-oblique slow mode has the small variation of $P_{tot}$\ in the
three-fluid plasmas consisting of fully-ionized hydrogen and a heavy ion
drifting along the background magnetic field.\textbf{\ }Therefore, one needs
addition tests for a more careful identification of the wave modes producing
the observed kinetic-scale PBSs.

In summary, we revealed several new mode properties of KAWs and KSWs
accounting for the kinetic effects of the ion and electron thermal pressure
and inertia. For KAWs, their frequency can reach and exceed the ion
cyclotron frequency at the ion kinetic scales, where both the thermal and
the inertial ion effects are important. The polarization properties of KAWs
are different in different $\beta $ ranges and depend on both the
propagation angle $\theta $ and the normalized perpendicular wavenumber $%
\rho k_{\perp }$. It appeared that KAWs undergo two reversals of electric
polarization defined by the zeros of denominator (\ref{x.tr}) and the
numerator (\ref{y.tr}) of (\ref{pol1}). In particular, in low-$\beta $
plasmas less oblique KAWs (\ref{theta2}) are left-hand polarized at longer
wavelengths and right-hand at shorter wavelengths. These properties are
important for the turbulence cascade transition across the ion-cyclotron
frequency, where it can be partially dissipated by the ion-cyclotron
resonance, in such a way that the left-hand KAWs possess stronger
dissipation as compared to the right-hand ones.

For KSWs, its frequency is always smaller than ion cyclotron frequency and
the mode is left-hand polarized. At the ion kinetic scales, $\rho k_{\perp
}\sim 1$, the electric and magnetic KSW components obey $|\delta E_{y}|\ll
|\delta E_{x}|$, $|\delta B_{x}|\ll |\delta B_{y}|$, and $|\delta
B_{z}|\simeq |\delta B_{y}|$. All these properties of KSWs can be described
approximately by the reduced expression obtained in the low-$\beta $ limit.

These new properties are important for understanding short-wavelength Alfv%
\'{e}n and slow modes and can be used in interpreting waves and turbulence
at kinetic scales.

\appendix

\section{The general dispersion equation}

From the momentum Equation (1), the ion and electron velocities are found as
\begin{eqnarray}
\Lambda _{i}\delta v_{ix} &=&-i\frac{\omega }{B_{0}\omega _{ci}}\delta E_{x}+%
\frac{1}{B_{0}}\delta E_{y}-\frac{\gamma T_{i}\omega k_{\perp }}{m_{i}\omega
_{ci}^{2}}\frac{\delta n}{n_{0}},  \label{A1} \\
\Lambda _{i}\delta v_{iy} &=&-\frac{1}{B_{0}}\delta E_{x}-i\frac{\omega }{%
B_{0}\omega _{ci}}\delta E_{y}+i\frac{\gamma T_{i}k_{\perp }}{m_{i}\omega
_{ci}}\frac{\delta n}{n_{0}},  \label{A2} \\
\delta v_{iz} &=&i\frac{e}{m_{i}\omega }\delta E_{z}+\frac{\gamma T_{i}k_{z}%
}{m_{i}\omega }\frac{\delta n}{n_{0}},  \label{A3} \\
\Lambda _{e}\delta v_{ex} &=&i\frac{Q\omega }{B_{0}\omega _{ci}}\delta E_{x}+%
\frac{1}{B_{0}}\delta E_{y}-\frac{Q\gamma T_{e}\omega k_{\perp }}{%
m_{i}\omega _{ci}^{2}}\frac{\delta n}{n_{0}},  \label{A4} \\
\Lambda _{e}\delta v_{ey} &=&-\frac{1}{B_{0}}\delta E_{x}+i\frac{Q\omega }{%
B_{0}\omega _{ci}}\delta E_{y}-i\frac{\gamma T_{e}k_{\perp }}{m_{i}\omega
_{ci}}\frac{\delta n}{n_{0}},  \label{A5} \\
\delta v_{ez} &=&-i\frac{e}{m_{e}\omega }\delta E_{z}+\frac{\gamma T_{e}k_{z}%
}{m_{e}\omega }\frac{\delta n}{n_{0}},  \label{A6}
\end{eqnarray}%
where $Q\equiv m_{e}/m_{i}$, $\Lambda _{i}\equiv 1-\omega ^{2}/\omega
_{ci}^{2}$, and $\Lambda _{e}\equiv 1-Q^{2}\omega ^{2}/\omega _{ci}^{2}$.
Note that the quasi-neutrality condition, $\delta n_{i}=\delta n_{e}\equiv
\delta n$, has been used in last derivation. By the use of expressions
(A1)--(A6), the current density $\delta \mathbf{J}=n_{0}e\left( \delta
\mathbf{v}_{i}-\delta \mathbf{v}_{e}\right) $ can be presented in the
following form:
\begin{eqnarray}
\Lambda _{i}\Lambda _{e}\delta J_{x} &=&-i\frac{n_{0}e\omega }{B_{0}\omega
_{ci}}\left( \Lambda _{e}+Q\Lambda _{i}\right) \delta E_{x}+\frac{n_{0}e}{%
B_{0}}\left( \Lambda _{e}-\Lambda _{i}\right) \delta E_{y}-\frac{\gamma
T\omega k_{\perp }}{B_{0}\omega _{ci}}\left( \Lambda _{e}\widetilde{T}%
_{i}-Q\Lambda _{i}\widetilde{T}_{e}\right) \delta n,  \label{A7} \\
\Lambda _{i}\Lambda _{e}\delta J_{y} &=&-\frac{n_{0}e}{B_{0}}\left( \Lambda
_{e}-\Lambda _{i}\right) \delta E_{x}-i\frac{n_{0}e\omega }{B_{0}\omega _{ci}%
}\left( \Lambda _{e}+Q\Lambda _{i}\right) \delta E_{y}+i\frac{\gamma
Tk_{\perp }}{B_{0}}\left( \Lambda _{e}\widetilde{T}_{i}+\Lambda _{i}%
\widetilde{T}_{e}\right) \delta n,  \label{A8} \\
\delta J_{z} &=&i\left( 1+Q\right) \frac{n_{0}e^{2}}{m_{e}\omega }\delta
E_{z}-\frac{e\gamma Tk_{z}}{m_{e}\omega }\left( \widetilde{T}_{e}-Q%
\widetilde{T}_{i}\right) \delta n,  \label{A9}
\end{eqnarray}%
where $T=T_{i}+T_{e}$ and $\widetilde{T}_{i,e}\equiv T_{i,e}/T$.

On the other hand, the current density can be expressed in terms of the
perturbed electric field only,
\begin{eqnarray}
\delta J_{x} &=&-i\frac{k_{z}^{2}}{\mu _{0}\omega }\delta E_{x}+i\frac{%
k_{\perp }k_{z}}{\mu _{0}\omega }\delta E_{z},  \label{A10} \\
\delta J_{y} &=&-i\frac{k^{2}}{\mu _{0}\omega }\delta E_{y},  \label{A11} \\
\delta J_{z} &=&i\frac{k_{\perp }k_{z}}{\mu _{0}\omega }\delta E_{x}-i\frac{%
k_{\perp }^{2}}{\mu _{0}\omega }\delta E_{z}.  \label{A12}
\end{eqnarray}

From Equations (A7) -- (A12), the electric field components can be expressed
in terms of $\delta n$:
\begin{eqnarray}
\Pi \delta E_{x} &=&ik_{\perp }\frac{\gamma T}{e}\Pi _{x}\frac{\delta n}{%
n_{0}},  \label{A13} \\
\Pi \delta E_{y} &=&k_{\perp }\frac{\gamma T}{e}\Pi _{y}\frac{\delta n}{n_{0}%
},  \label{A14} \\
\Pi \delta E_{z} &=&ik_{z}\frac{\gamma T}{e}\Pi _{z}\frac{\delta n}{n_{0}},
\label{A15}
\end{eqnarray}%
where
\begin{eqnarray}
\Pi &=&\left( 1+Q\right) ^{2}\left( 1+Q+\lambda _{e}^{2}k_{\perp
}^{2}\right) \omega ^{4}  \notag \\
&&-\left( 1+Q\right) \left[ 1+Q+\lambda _{e}^{2}k_{\perp }^{2}+\left(
1+Q\right) k_{z}^{2}/k^{2}\right] \Lambda _{h}V_{A}^{2}k^{2}\omega ^{2}
\notag \\
&&+\left( 1+Q\right) \Lambda _{i}\Lambda _{e}V_{A}^{4}k^{2}k_{z}^{2},  \notag
\\
\Pi _{x} &=&\left( 1+Q\right) \left( 1+Q+\lambda _{e}^{2}k_{\perp
}^{2}\right) \left( Q\widetilde{T}_{i}-\widetilde{T}_{e}\right) \omega ^{4}
\notag \\
&&-\left[ \left( 1+Q+\lambda _{e}^{2}k_{\perp }^{2}\right) \left( \Lambda
_{e}\widetilde{T}_{i}-Q\Lambda _{i}\widetilde{T}_{e}\right) +\left(
1+Q\right) \Lambda _{h}\left( Q\widetilde{T}_{i}-\widetilde{T}_{e}\right)
k_{z}^{2}/k^{2}\right] V_{A}^{2}k^{2}\omega ^{2}  \notag \\
&&+\Lambda _{i}\Lambda _{e}\left( Q\widetilde{T}_{i}-\widetilde{T}%
_{e}\right) V_{A}^{4}k^{2}k_{z}^{2},  \notag \\
\Pi _{y} &=&\left( 1+Q\right) \left[ \left( 1+Q+\lambda _{e}^{2}k_{\perp
}^{2}\right) \omega ^{2}-\Lambda _{h}V_{A}^{2}k_{z}^{2}\right] \omega \omega
_{ci}^{{}},  \notag \\
\Pi _{z} &=&\Pi _{x}+\left( 1-Q^{2}\right) V_{A}^{2}k^{2}\omega ^{2},  \notag
\end{eqnarray}%
and $\Lambda _{h}\equiv \left( 1-Q\omega ^{2}/\omega _{ci}^{2}\right) $.

Now we can use expressions (A13)-(A15) to eliminate the electric field from
the number density equation
\begin{equation}
\left[ \left( \Lambda _{i}+\rho _{i}^{2}k_{\perp }^{2}\right) \omega
^{2}-\Lambda _{i}V_{Ti}^{2}k_{z}^{2}\right] \frac{\delta n}{n_{0}}=-i\frac{%
\omega ^{2}k_{\perp }}{B_{0}\omega _{ci}}\delta E_{x}+\frac{\omega k_{\perp }%
}{B_{0}}\delta E_{y}+i\frac{e}{m_{i}}k_{z}\Lambda _{i}\delta E_{z},
\label{A16}
\end{equation}%
which results in the general dispersion equation:
\begin{eqnarray}
&&\omega ^{6}\left( 1+Q\right) \left( 1+Q+\lambda _{e}^{2}k^{2}\right) ^{2}
\notag \\
&&-\omega ^{4}\left[ \left( 1+Q\right) \left( 1+Q+\lambda
_{e}^{2}k^{2}\right) +\left( 1+Q+\lambda _{e}^{2}k^{2}\right)
^{2}V_{T}^{2}/V_{A}^{2}+\left( 1+Q^{3}\right) \lambda
_{i}^{2}k_{z}^{2}+\left( 1+Q\right) ^{2}k_{z}^{2}/k^{2}\right] V_{A}^{2}k^{2}
\notag \\
&&+\omega ^{2}\left[ \left( 1+Q\right) \left( 1+2V_{T}^{2}/V_{A}^{2}\right)
+\left( 1+Q^{2}\right) \rho ^{2}k^{2}\right] V_{A}^{4}k^{2}k_{z}^{2}  \notag
\\
&&-\beta V_{A}^{6}k^{2}k_{z}^{4}=0,  \label{A17}
\end{eqnarray}%
where $V_{T}=\sqrt{\gamma T/m_{i}}$, $\rho ^{2}=\rho _{i}^{2}+\rho _{s}^{2}$%
, $\rho _{i}$ is the ion gyroradius, $\rho _{s}$ is the ion-acoustic
gyroradius, and $\lambda _{i}$ is the ion inertial length.

In deriving Equation (A17), we neglected the displacement current in the
Ampere's law, but kept all other terms. Also, we treated the electrons and
ions separately. This makes our derivation and results different from the
derivations by Stringer (1963) and Bellan (2012). The above two authors used
equations of the mass motion and the generalized Ohm's law (Equations (A1)
and (A2) in Stringer (1963)) with one-fluid variables $\rho
=\sum\limits_{\alpha =i,e}m_{\alpha }n_{\alpha }$\ and $v=\sum\limits_{%
\alpha =i,e}m_{\alpha }n_{\alpha }v_{\alpha }/\sum\limits_{\alpha
=i,e}m_{\alpha }n_{\alpha }$\ where some terms of order $Q$\ were discarded.
The resulting general dispersion equation is (Stringer, 1963)\textbf{\ }%
\begin{eqnarray}
&&\omega ^{6}\left( 1+\lambda _{e}^{2}k^{2}\right) ^{2}  \notag \\
&&-\omega ^{4}\left[ \left( 1+\lambda _{e}^{2}k^{2}\right) +\left( 1+\lambda
_{e}^{2}k^{2}\right) ^{2}V_{T}^{2}/V_{A}^{2}+\left( 1+Q\right) \lambda
_{i}^{2}k_{z}^{2}+k_{z}^{2}/k^{2}\right] V_{A}^{2}k^{2}  \notag \\
&&+\omega ^{2}\left[ \left( 1+2V_{T}^{2}/V_{A}^{2}\right) +\left(
1+2Q\right) \rho ^{2}k^{2}\right] V_{A}^{4}k^{2}k_{z}^{2}  \notag \\
&&-\beta V_{A}^{6}k^{2}k_{z}^{4}=0.  \label{A18}
\end{eqnarray}%
Some terms in our expression (A17) and in Stringer's expression (A18) are
different. The differences come from the different treatment of some minor
terms $\sim Q$: all small terms are kept in our derivation but an incomplete
set of terms was used by Stringer (1963). Since the major terms in Equations
(A17) and (A18) are the same, the resulting dispersion relations for the
fast, Alfv\'{e}n and slow modes are also nearly the same. However, behavior
of some polarization ratios differ significantly.

\section{Polarization and Correlation}

The wave properties involving wave polarization and correlation are
summarized in \cite{KR94}, here we repeat these definitions for convenient
discussion. The electric field polarization with respect to the ambient
magnetic-field is defined as
\begin{equation}
P_{E,\mathbf{B_{0}}}=\frac{\delta E_{y}}{i\delta E_{x}}.  \label{B1}
\end{equation}%
The right-hand polarized mode corresponds to $\mathrm{Re}\left( P_{E,\mathbf{%
B}_{0}}\right) >0$, and left-hand polarized mode corresponds to $\mathrm{Re}%
\left( P_{E,\mathbf{B}_{0}}\right) <0$. $\mathrm{Rm}\left( P_{E,\mathbf{B}%
_{0}}\right) =\pm 1$ correspond to the right- or left-hand circularly
polarized mode. Note that the definition $P_{E,\mathbf{B_{0}}}=i\delta
E_{x}/\delta Ey$ is used in \cite{KR94}.

The magnetic field polarization with respect to the ambient magnetic-field
is defined as
\begin{equation}
P_{B,\mathbf{B_{0}}}=\frac{\delta B_{y}}{i\delta B_{x}},  \label{B2}
\end{equation}%
and the magnetic field polarization with respect to wave vector is
\begin{subequations}
\begin{equation}
P_{B,\mathbf{k}}=\frac{i\delta \mathbf{B}\cdot \left( \hat{\mathbf{z}}\times
\hat{\mathbf{e}}_{y}\right) }{\delta B_{y}}=\frac{1}{P_{\delta B,\mathbf{%
B_{0}}}\mathrm{cos}\theta }.  \label{B3}
\end{equation}

The magnetic helicity is expressed as
\end{subequations}
\begin{equation}
\sigma =\frac{k\left( \mathbf{A}\cdot \delta \mathbf{B}^{\ast }\right) }{%
\delta B^{2}}=\frac{2\mathrm{Re}\left( P_{B,\mathbf{k}}\right) }{%
1+\left\vert P_{B,\mathbf{k}}\right\vert ^{2}},  \label{B4}
\end{equation}%
where positive or negative helicity corresponds to a left- or right-hand
sense of rotation with respect to $\mathbf{k}$ \cite{GA86}, respectively.

The magnetic field $-$ density correlation corresponds to
\begin{equation}
C_{\parallel }=\frac{\delta n/n_{0}}{\delta B_{z}/B_{0}},  \label{B5}
\end{equation}%
Correspondingly, the thermal pressure $-$ magnetic pressure correlation is
defined as
\begin{equation}
C_{PBn}=\frac{\gamma \left( T_{i}+T_{e}\right) \delta n}{\delta
B_{z}B_{0}/\mu _{0}}=\beta C_{\parallel },  \label{B6}
\end{equation}%
hence, the total pressure perturbation is written as $\delta P_{tot}=\left(
1+\beta C_{\parallel }\right) {\delta B_{z}B_{0}/\mu _{0}}$.

Compressibility
\begin{equation}
C_{Bn}=\frac{\delta n^{2}/n_{0}^{2}}{\delta B^{2}/B_{0}^{2}}=\mathrm{sin}%
^{2}\theta \left\vert C_{\parallel }\right\vert ^{2}\frac{\left\vert
P_{\delta B,\mathbf{k}}\right\vert ^{2}}{1+\left\vert P_{\delta B,\mathbf{k}%
}\right\vert ^{2}},  \label{B7}
\end{equation}%
describes the relation between the total magnetic field perturbation and the
number density.

The Alfv\'{e}n ratio and cross helicity for $j$ species are defined as
\begin{eqnarray}
&&R_{A}^{j}=\mu _{0}n_{0}m_{i}\frac{\left\vert \delta \mathbf{v}%
_{j}\right\vert ^{2}}{\left\vert \delta \mathbf{B}\right\vert ^{2}},
\label{B8} \\
&&\sigma _{C}^{j}=2\frac{\left( \mu _{0}n_{0}m_{i}\right) ^{1/2}\mathrm{Re}%
\left( \delta \mathbf{v}_{j}\cdot \delta \mathbf{B}^{\ast }\right) }{\mu
_{0}n_{0}m_{i}\left\vert \delta \mathbf{v}_{j}\right\vert ^{2}+\left\vert
\delta \mathbf{B}\right\vert ^{2}},  \label{B9}
\end{eqnarray}%
which gives the correlation between the perturbed velocity and magnetic
field.

\section{Linear dispersion and wave parameters in the low-$\protect\beta $
plasma}

In the low-beta plasma, $\beta \ll 1$, the linear dispersion and relations
among field and plasma quantities are simpler than (6)--(7). For KAWs, the
dispersion relation
\begin{equation}
\omega ^{2}=V_{A}^{2}k_{z}^{2}\mathcal{R}/\mathcal{L}^{\prime },  \label{C1}
\end{equation}%
and 
\begin{eqnarray}
\frac{\delta \mathbf{E}}{V_{A}} &=&\frac{\mathcal{R}_{i}\mathcal{L}-\rho
_{s}^{2}k_{\perp }^{2}\lambda _{i}^{2}k_{z}^{2}}{\mathcal{R}^{1/2}\mathcal{L}%
^{\prime 1/2}}\delta B_{y}\hat{\mathbf{e}}_{x}+i\frac{k_{z}}{k_{\perp }^{{}}}%
\frac{\left( \mathcal{L}\rho ^{2}k_{\perp }^{2}-\lambda
_{i}^{2}k_{z}^{2}\right) }{\lambda _{i}k_{\perp }^{{}}\mathcal{L}^{\prime }}%
\delta B_{y}\hat{\mathbf{e}}_{y}-\frac{k_{z}}{k_{\perp }}\frac{\left(
1+\lambda _{i}^{2}k_{z}^{2}\right) \rho _{s}^{2}k_{\perp }^{2}-\mathcal{R}%
_{i}\lambda _{e}^{2}k_{\perp }^{2}}{\mathcal{R}^{1/2}\mathcal{L}^{\prime 1/2}%
}\delta B_{y}\hat{\mathbf{e}}_{z},  \notag \\
\delta \mathbf{B} &=&-i\frac{k_{z}}{k_{\perp }^{{}}}\frac{\left( \mathcal{L}%
\rho ^{2}k_{\perp }^{2}-\lambda _{i}^{2}k_{z}^{2}\right) }{\lambda
_{i}k_{\perp }^{{}}\mathcal{R}^{1/2}\mathcal{L}^{\prime 1/2}}\delta B_{y}%
\hat{\mathbf{e}}_{x}+\delta B_{y}\hat{\mathbf{e}}_{y}+i\frac{\mathcal{L}\rho
^{2}k_{\perp }^{2}-\lambda _{i}^{2}k_{z}^{2}}{\lambda _{i}k_{\perp }\mathcal{%
R}^{1/2}\mathcal{L}^{\prime 1/2}}\delta B_{y}\hat{\mathbf{e}}_{z},  \notag \\
\frac{\delta \mathbf{v}_{i}}{V_{A}} &=&-i\lambda _{i}k_{z}\frac{\delta B_{y}%
}{B_{0}}\hat{\mathbf{e}}_{x}-\frac{\mathcal{L}^{\prime 1/2}}{\mathcal{R}%
^{1/2}}\frac{\delta B_{y}}{B_{0}}\hat{\mathbf{e}}_{y}-i\frac{\left( \mathcal{%
R}-\mathcal{L}+\rho ^{2}k_{\perp }^{2}\lambda _{i}^{2}k_{z}^{2}\right) }{%
\lambda _{i}k_{\perp }\mathcal{R}}\frac{\delta B_{y}}{B_{0}}\hat{\mathbf{e}}%
_{z},  \notag \\
\frac{\delta \mathbf{v}_{e}}{V_{A}} &=&i\frac{k_{z}}{k_{\perp }^{{}}}\frac{%
\left( \mathcal{L}\rho ^{2}k_{\perp }^{2}-\lambda _{i}^{2}k_{z}^{2}\right)
\mathcal{R}}{\lambda _{i}k_{\perp }^{{}}\mathcal{L}^{\prime }}\frac{\delta
B_{y}}{B_{0}}\hat{\mathbf{e}}_{x}-\frac{\mathcal{L}\mathcal{R}^{1/2}}{%
\mathcal{L}^{\prime 1/2}}\frac{\delta B_{y}}{B_{0}}\hat{\mathbf{e}}%
_{y}-i\lambda _{i}k_{\perp }\frac{\delta B_{y}}{B_{0}}\hat{\mathbf{e}}_{z},
\label{C2}
\end{eqnarray}%
where
\begin{equation*}
\mathcal{R}=1+\rho ^{2}k_{\perp }^{2},~~\mathcal{R}_{i}=1+\rho
_{i}^{2}k_{\perp }^{2},~~\mathcal{L}=1+\lambda _{e}^{2}k_{\perp }^{2},~~%
\mathcal{L}^{\prime }=1+\lambda _{e}^{2}k_{\perp }^{2}+\lambda
_{i}^{2}k_{z}^{2}.
\end{equation*}

For KSWs, the dispersion relation
\begin{equation}
\omega ^{2}=V_{T}^{2}k_{z}^{2}/\mathcal{R},  \label{C3}
\end{equation}%
and
\begin{eqnarray}
&&\frac{\delta \mathbf{E}}{V_{T}}=\frac{\mathcal{L}\widetilde{T}_{i}+%
\mathcal{R}\widetilde{T}_{e}/\beta }{\mathcal{R}^{1/2}}\delta B_{y}\hat{%
\mathbf{e}}_{x}-i\frac{k_{z}}{\rho k_{\perp }^{2}}\delta B_{y}\hat{\mathbf{e}%
}_{y}+\frac{k_{z}}{k_{\perp }}\frac{\mathcal{L}\widetilde{T}_{i}+\mathcal{R}%
\widetilde{T}_{e}/\beta -1}{\mathcal{R}^{1/2}}\delta B_{y}\hat{\mathbf{e}}%
_{z},  \notag \\
&&\delta \mathbf{B}=i\frac{k_{z}}{\rho k_{\perp }^{2}}\mathcal{R}%
^{1/2}\delta B_{y}\hat{\mathbf{e}}_{x}+\delta B_{y}\hat{\mathbf{e}}_{y}-i%
\frac{\mathcal{R}^{1/2}}{\rho k_{\perp }}\delta B_{y}\hat{\mathbf{e}}_{z},
\notag \\
&&\frac{\delta \mathbf{v}_{i}}{V_{T}}=-i\frac{k_{z}\left( 1+\lambda
_{i}^{2}k_{\perp }^{2}\right) }{\rho k_{\perp }^{2}}\frac{\delta B_{y}}{B_{0}%
}\hat{\mathbf{e}}_{x}-\frac{\mathcal{R}^{1/2}}{\beta }\frac{\delta B_{y}}{%
B_{0}}\hat{\mathbf{e}}_{y}+i\frac{\mathcal{R}}{\beta \rho k_{\perp }}\frac{%
\delta B_{y}}{B_{0}}\hat{\mathbf{e}}_{z},  \notag \\
&&\frac{\delta \mathbf{v}_{e}}{V_{T}}=-i\frac{k_{z}}{\rho k_{\perp }^{2}}%
\frac{\delta B_{y}}{B_{0}}\hat{\mathbf{e}}_{x}-\frac{\mathcal{L}}{\mathcal{R}%
^{1/2}}\frac{\delta B_{y}}{B_{0}}\hat{\mathbf{e}}_{y}+i\frac{1}{\beta \rho
k_{\perp }}\frac{\delta B_{y}}{B_{0}}\hat{\mathbf{e}}_{z}  \label{C4}
\end{eqnarray}


%

\begin{acknowledgments}
This research was supported by the Belgian Federal Science Policy Office via
IAP Programme (project P7/08 CHARM), by the European Commission via FP7
Program (project 313038 STORM), by NSFC under grant No. 11303099, No.11373070, and No. 41074107, by MSTC
under grant No. 2011CB811402, by NSF of Jiangsu Province under grant No.
BK2012495, and by Key Laboratory of Solar Activity at NAO, CAS, under grant
No. KLSA201304.

\end{acknowledgments}

%
%
%
%
%
%
%
%
%


\end{document}